\documentstyle[prl,aps,multicol,epsf]{revtex}
\begin{document}
\widetext
\draft

\title{Magnetic reconstruction at (001) CaMnO$_3$ surface}
\author{Alessio Filippetti and Warren E. Pickett}
\address{Department of Physics, University of California at Davis, 
Davis, California 95616}
\maketitle
\begin{abstract}
  The Mn-terminated (001) surface of the stable anti-ferromagnetic 
  insulating phase of cubic perovskite CaMnO$_3$ is found to undergo a 
  magnetic reconstruction consisting on a spin-flip process at surface:
  each Mn spin at the surface flips to pair with that of Mn in the 
  subsurface layer.
  In spite of very little Mn-O charge transfer at surface, the surface 
  behavior is driven by the $e_g$ states due to $d_{xy}$ $\rightarrow$
  $d_{z^2}$ charge redistribution.
  These results, based on local spin density theory, give a 
  double exchange like coupling that is driven by $e_g$ character, not
  additional charge, and may have relevance to CMR materials.

\end{abstract}

\pacs{73.20.At,75.30.Pd,75.30.Vn,75.25.+z}

\begin{multicols}{2}

Despite the abundance of work on manganese-based perovskites in the
attempt to understand the rich panorama of their bulk 
properties,\cite{book,ps} very little
is known about their surfaces\cite{surf,surf1,surf2} or interfaces.
The physical mechanism inducing the so-called colossal magnetoresistance 
(CMR) in La$_{1-x}$D$_x$MnO$_3$ (with D a divalent alkaline
earth ion, and $x \sim 0.3$) 
is yet to be fully understood, although it seems clear that the almost 
half-metallic nature\cite{ps}
(i.e. the complete spin-polarization of the electrons at the Fermi level)
plays a decisive role.  The clearest evidence so far of half-metallicity
is from photoemission spectra,\cite{surf} whose surface sensitivity makes it
essential to know the electronic structure of the surface itself.
The surface introduces the likelihood of square pyramidal coordinated
Mn, which is also essential to the understanding of oxygen-deficient
perovskite manganites.\cite{Zampieri}  In addition, the interfacial
behavior that is critical in producing low field CMR in polycrystalline
material\cite{poli} and trilayer junctions\cite{tri} will involve
closely related effects due to symmetry lowering of the Mn ion.
Thus, studies of the surfaces of manganites are timely.

The electronic structure of the stable phase of bulk CaMnO$_3$ has been 
actively investigated in recent years\cite{Zampieri,ps,other}. 
CaMnO$_3$ is a G-type 
antiferromagnetic (AFM) semiconductor.
The nominal ionic picture Ca$^{2+}$Mn$^{4+}$O$^{2-}_3$ with spherical 
Mn d$^3$ configuration, makes the cubic (fcc) phase stable\cite{stable} 
over possible distortions observed for instance in LaMnO$_3$.
In the G-type arrangement all nearest neighbors in the simple-cubic 
sublattice of Mn have spin-antiparallel orientation. The chemical picture 
of Mn$^{4+}$ ions is represented by completely occupied 
d t$_{2g}^{\uparrow}$ states. An energy gap of $\sim$ 0.4 eV separate them
from the empty d e$_g^{\uparrow}$ orbitals. Hybridization with
O p states reduces the 3 $\mu_B$ nominal magnetization of Mn to 
$\sim$ 2.5 $\mu_B$, whereas magnetic moments on O or Ca are zero by symmetry.

In this paper we study the simplest manganite surface, the
(001) surface of cubic CaMnO$_3$, to determine the
surface-induced changes of structural, electronic, and magnetic properties.
We find unexpectedly rich effects of surface symmetry lowering:
a spin-flip occurs on the surface Mn ions that can be traced
to surface states that redistribute charge and spin amongst the various
Mn $d$ suborbitals and render it metallic without doping.  
The net effect is a short-range double-exchange-like
phenomenon that relates metallicity and spin alignment, analogous to
the CMR phases.

Calculations were done in a local-spin-density framework; the 
exchange-correlation potential formula by Perdew and Zunger\cite{pz} 
was used. A plane-wave basis with 30 Ryd cut-off energy and Vanderbilt 
pseudopotentials\cite{van} make the computation viable.
To establish the accuracy of our methods, which have only recently been
applied to magnetic materials,
in Table \ref{bulk} we report our results for the bulk CaMnO$_3$ in different
magnetic phases. As already shown, for manganese perovskites the local 
spin density approximation successfully predicts the observed stable 
phase not only against the strongly unfavoured paramagnetic (PM) phase, but 
also in competition with closer configurations like the ferromagnetic (FM) and 
the A-type AFM (made of (001) FM layers alternating along [001] direction).
Our results are in very good agreement with those of previous all-electron 
linear-augmented plane-wave calculations\cite{ps}. 
Also, our calculated value for the equilibrium lattice constant 
of G-type AFM phase (3.735 \AA) is in almost perfect agreement with the 
experimental value (3.729 \AA). 

\begin{figure}
\epsfclipon
\epsfxsize=5cm
\centerline{\epsffile{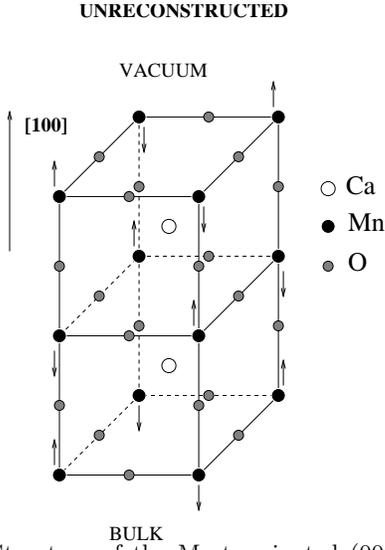}}
\caption{Structure of the Mn-terminated (001) surface of G-type AFM CaMnO$_3$
in the unreconstructed magnetic configuration, i.e. the spin orientation at
surface is equal to that in the bulk.
\label{fig1}}
\end{figure}

The stacking along [001] consists of alternating MnO$_2$ and CaO units 
(see Fig. \ref{fig1}), and the surface unit cell of G-type AFM CaMnO$_3$ 
is $\sqrt{2}\times\sqrt{2}$ with respect to that 
of the bulk cubic cell (we neglect the very small structural 
distortion\cite{stable}).  The (001) layers are individually AFM and neutral,
so the surface is formally non-polar. 
Surface formation produces two different 
surfaces, i.e. Mn-terminated and Ca-terminated.
The presence of two inequivalent surfaces in a slab will produce fictitious 
fields in the vacuum that could affect the electronic and magnetic structure
at the surface.  We are interested on the Mn-terminated surface, 
since on it the effects on magnetic properties due to the surface 
formation are most visible. Thus we use a slab containing 
two identical Mn-terminated surfaces, with mirror symmetry 
in the central Mn layer (in total a 46-atom slab with 9 layers of atoms and 
3 of vacuum). 

Surface neutrality generally favours
the stability of the ideal surface against reconstructions involving strong
changes of symmetry and atomic density at surface.
Therefore in this work we consider the surface with relaxed but structurally
unreconstructed structure, with different types of magnetic order. 
Thus we will speak of `reconstruction' in purely magnetic sense:
on the unreconstructed surface the spins are oriented as in the bulk 
(Figure \ref{fig1}), while the reconstructions involve spin-flips 
on the surface layer.

\begin{figure}
\epsfclipon
\epsfxsize=8cm
\epsffile{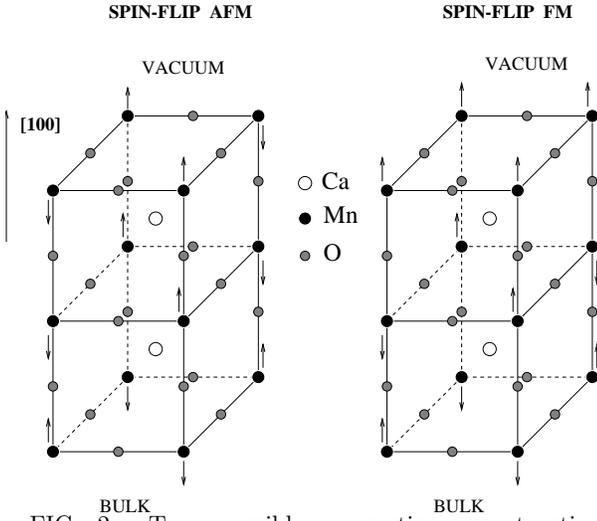}
\caption{Two possible magnetic reconstructions of Mn-terminated (001) surface.
Left picture: spins on both Mn at surface are flipped, the surface is still
AFM. Right picture: spin on one of two Mn is flipped: FM surface.
\label{fig2}}
\end{figure}

The structure of the
two configurations that can be obtained by flipping surface spins 
are pictured in Fig. \ref{fig2}.
In the left panel all surface spins are flipped, so vectors $(\pm a,\pm a)$
remain AFM translations but each surface spin is aligned with its
subsurface neighbor (spin-flip AFM: sf-AFM).
In the right panel only one (of two) surface spins is flipped, leaving 
a FM surface layer (spin-flip FM: sf-FM).

Magnetic and relaxation energies and workfunctions for the three 
phases are reported in Table \ref{tavola}. 
The $\Delta E$'s reported in Table \ref{tavola} are the energies gained by
relaxing all the atoms into the slab from their ideal positions. They are
small and reflect the small inward atomic displacements
$\sim$ 1\% of the cubic lattice constant.
This indicates a low excess stress due to the surface
formation, and suggests that structural reconstructions are unlikely.
The workfunction depends very little on the spin arrangement
but is largest for the most stable surface.
Most significantly, the sf-AFM surface is stable against
the unreconstructed one, whereas the sf-FM is the most unfavoured.
Thus, a quite intriguing physical picture follows: at the surface each 
spin prefers to pair with the one in the subsurface layer, while still keeping
the AFM arrangement in-plane. 
 
It is possible to express the energy differences for 
differing types of magnetic 
order in terms of exchange constants in a Heisenberg model 
\begin{equation}
H = - \sum_{<ij>} J_{ij} \hat S_i \cdot \hat S_j
\end{equation}
where $\hat S_j$ is a unit vector in the direction of the moment on site
$j$, and the sum is over distinct pairs.
For the bulk we get first and second neighbor constants $J_1=-26$ meV,
$J_2=-4$ meV.  This small value of $J_2$ suggest that the nearest neighbor
(nn) exchange constants contain the important contributions. 
From surface energies we get the 
nn coupling parallel and normal to the surface of
$J^{\parallel}$= -22 meV, $J^{\perp}$ = 29 meV.  While $J^{\parallel}$
is close to the bulk value, $J^{\perp}$ has the opposite sign and
is larger in magnitude,
indicating the FM alignment of surface and subsurface spins is robust.

\begin{figure}
\epsfclipon
\epsfxsize=8cm
\epsffile{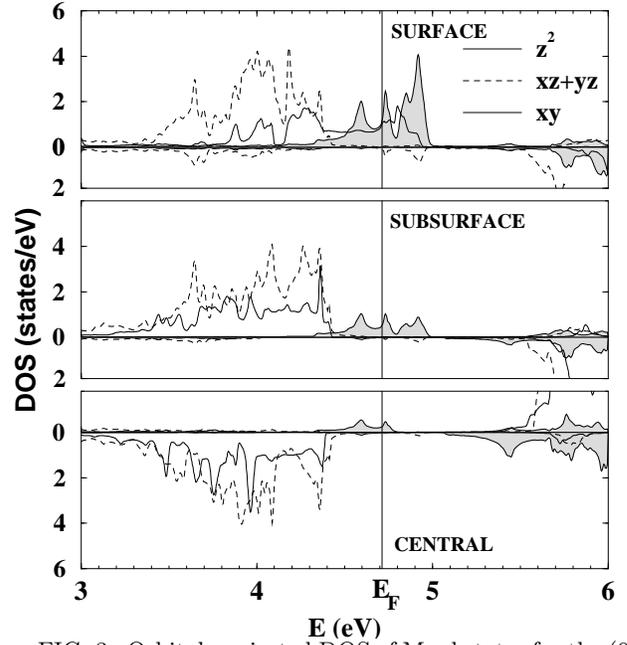}
\caption{Orbital-projected DOS of Mn d states for the (001) spin-flip AFM
surface. Upper, middle and lower panels refer to surface, sub-surface and
central Mn atoms in the slab, respectively. Dashed, solid, and thick
solid lines refer to $d_{xz}$+$d_{yz}$, $d_{z^2}$, and $d_{xy}$ orbitals,
respectively.
Up spin  DOS is plotted upward; there is another atom in each layer
whose DOS is flipped from those shown.
\label{dos}}
\end{figure}

The reversal of the surface-subsurface coupling
can be traced to redistribution of $d$ suborbital 
occupations compared to the bulk, due to the occurrence of surface states.
The orbital-projected Mn 4$d$ density of states 
(DOS) of the stable sf-AFM phase near the Fermi level, shown 
in Fig. \ref{dos}, 
makes evident the surface states that lie within the bulk band gap, which
extends from -0.3 eV to 0.1 eV relative to E$_F$.
The surface states are of two distinct types, $d_{z^2}$ and $d_{xy}$,
reflecting
the strong symmetry lowering of both $e_g \rightarrow d_{z^2}, d_{x^2-y^2}$ 
and $t_{2g} \rightarrow d_{xy},(d_{xz},d_{yz})$ manifolds.
The surface states are almost 
completely
polarized, a result of the large spin splitting $\Delta_{ex}$
=2 eV that strongly inhibits hopping between ions of different spin.

\begin{figure}
\epsfclipon
\epsfxsize=9cm
\centerline{\epsffile{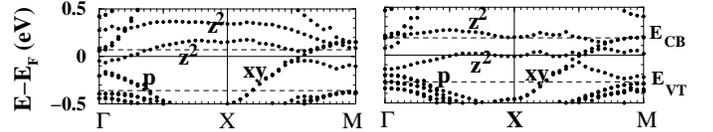}}
\caption{Band structure in a small energy window around the Fermi energy.
Left panel refers to the unreconstructed surface, right panel to the
spin-flip AFM. Dashed lines are boundaries of the bulk energy gap.
High symmetry points in the IBZ are $\Gamma$=(0,0,0), X=(2$\pi$/a')(1/2,0,0),
M=(2$\pi$/a')(1/2,1/2,0), in the $\sqrt{2}\times\sqrt{2}$ surface cell.}
\label{band}
\end{figure}

Figure \ref{band} presents the 
surface band structure, where for clarity, only the energy region of interest
(roughly spanning the bulk gap) is shown. The $d_{xy}$ and $d_{z^2}$
surface states are easily identifiable. 
The $d_{xy}$ states at the surface are shifted upward by 1 eV and
overlap with $d_{z^2}$ states that in the bulk hybridize strongly with the
O $p\sigma$ orbitals and form low-lying bonding and high-lying antibonding
bands (most of the d weight is in the latter).
The two $d_{z^2}$ states
(one from each surface of the slab) are split by 0.2 eV as a
result of the interaction between Mn at opposite sides of the slab.
(For a thicker slab they would converge to a single, doubly degenerate band,
averaging the calculated $d_{z^2}$ states).
The $d_{xy}$ band has a bandwidth of 1.4 eV and a dispersion that follows
\begin{equation}
\varepsilon_k^{xy} = -2t [cos(k_x+k_y) a + cos(k_x-k_y) a]
\end{equation}
using the conventional perovskite coordinates.  This dispersion arises from
hopping between second nn, which are the nearest neighbors of like spin.
The effective
hopping amplitude is $t$ = 0.17 eV.  The $d_{z^2}$ band is very narrow
(0.2 eV) and its dispersion is not easily represented by tight binding
form, reflecting small competing hopping processes along the
surface (and perhaps subsurface) that are not 
easily identified.  Coupling of the $d_{z^2}$ state {\it perpendicular} to the
surface is large, however, as reflected in the penetration of the state
onto the fifth atomic layer (third Mn layer).

The net effect on the Mn ion of surface formation is an 
intra-atomic shift of charge from $d_{xy}$ to $d_{z^2}$ orbital, 
without appreciable change of charge or moment.
In the solid the magnetic moment comes 
almost entirely from t$_{2g}$ states, whereas on the surface 
$d_{z^2}$ (surface) states contribute about 30\% of the moment. 
Only the subsurface Mn in the 
sf-AFM phase experiences a net gain (0.07 $\mu_B$) due to the partially 
occupied $d_{z^2}$ state not compensated by a depletion of $d_{xy}$ states.
In bulk, magnetic moments in the G-type bulk are only allowed on Mn, 
by symmetry. With the surface formation, O in-plane with Ca 
acquire a magnetic moment as well. This is larger in the sf-AFM 
(0.11 $\mu_B$) than in the unreconstructed phase (0.06 $\mu_B$), because it is 
enhanced by the parallel alignment of two neighboring Mn spins.

\begin{figure}
\epsfclipon
\epsfxsize=8cm
\epsffile{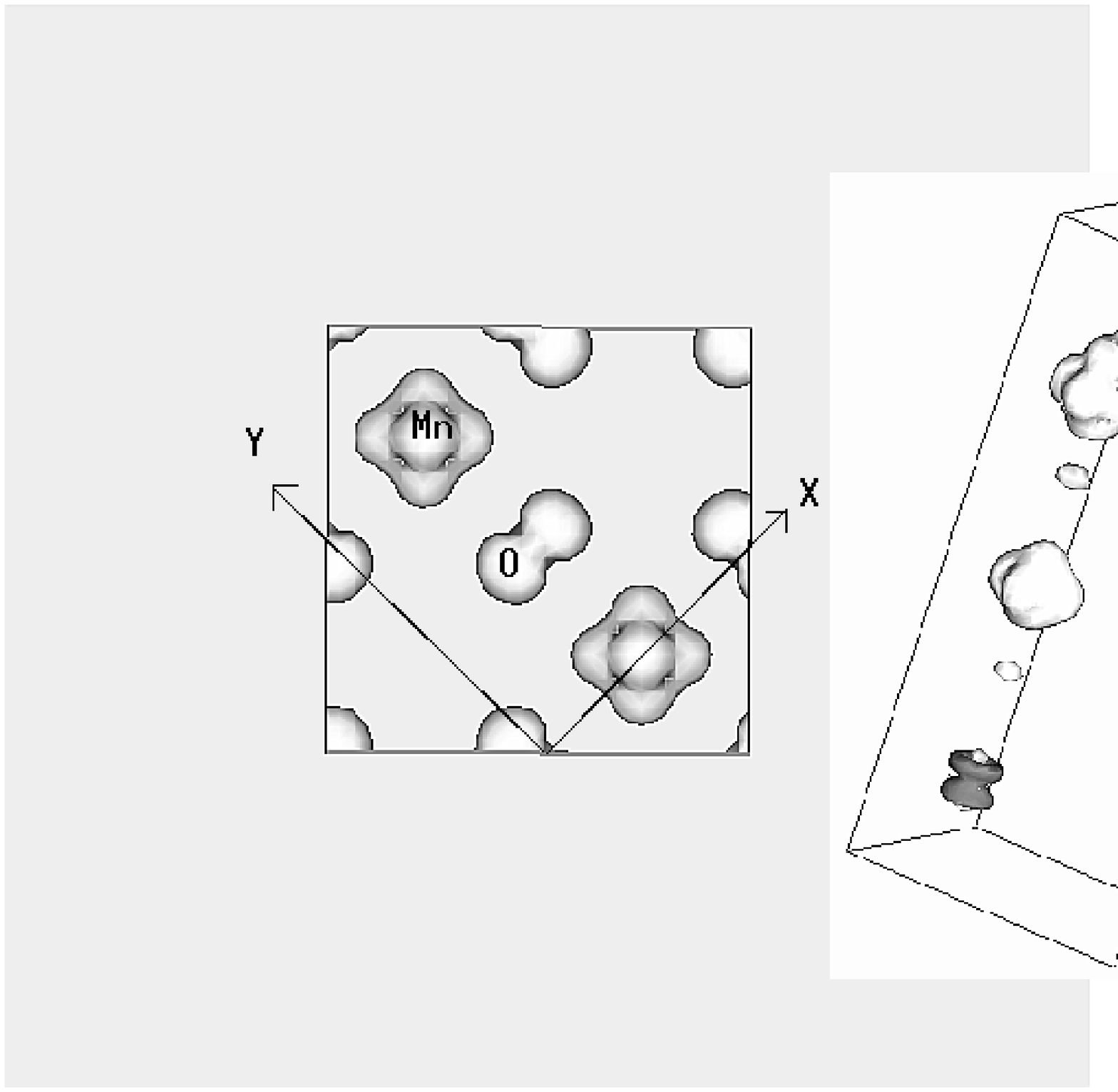}
\caption{Isosurface plots of the surface states for the spin-flip
AFM phase. Left panel: top view of the electron charge isosurface 
(of 0.01 electron/Bohr$^{-3}$ magnitude) at (100) surface. 
Right panel: tri-dimensional view of the magnetization. By inversion 
symmetry along $\hat{z}$ only half slab (five layers) is shown; 
on top there is the vacuum, on bottom the bulk. 
Light and dark isosurfaces are of same magnitude (0.005 $\mu_B$/Bohr$^{-3}$)
and opposite sign.
\label{isos}}
\end{figure}

Most of the characteristics inferred by DOS and band structure
analysis can be better visualized by means of isosurface plots 
(Fig. \ref{isos}) of charge density and magnetization
of the stable sf-AFM phase. 
The quantities shown are due only to states within the 
bulk gap (see Figure \ref{band}), thus represent charge and 
magnetization of the surface states. The charge density 
clearly shows both $d_{xy}$ and $d_{z^2}$ characters of the charge on Mn,
as well as a $p_{\pi}$-type contribution from O.
 
The physical mechanism driving the changes in exchange
interaction parameters at the surface is related to that described by
Solovyev {\it et al.}\cite{orbo} investigating how the Jahn-Teller 
distortions (JTD) affect the magnetic ordering of LaMnO$_3$.
The basic driving force in the FM-to-AFM transition of LaMnO$_3$ vs. JTD
is the decreasing of $d_{z^2}$ occupancy occurring with JTD. The 
$d_{z^2}-d_{z^2}$
interaction is indeed a dominant positive (i.e. FM) contribution 
to $J^{\perp}$. 
Also positive are the $d_{x^2-y^2}-d_{z^2}$ and the much weaker 
$d_{x^2-y^2}-d_{x^2-y^2}$ interactions, whereas t$_{2g}$ orbitals
interact by superexchange, and their contribution is AFM\cite{orbo}. 
When the $d_{z^2}$ orbitals are sufficiently occupied to make $J^{\perp}$ 
larger than the next nearest neighbor interactions (favouring AFM) the 
order becomes FM along the $\hat{z}$ axis.
 
In the present case the picture follows analogously: surface formation,
not JTD, results in dehybridization and
partial filling of the $d_{z^2}$ states on surface 
(and subsurface) Mn, and a partial depletion of $d_{xy}$ orbitals.  
As a consequence $J^{\perp}$ 
changes sign and the magnetic ordering along $\hat{z}$ reverses.
A verification of this mechanism is given by comparison of the
band structures (or DOS) of the two competing phases (Figure \ref{band}): 
in the sf-AFM phase there is more $d_{z^2}$ occupation and less $d_{xy}$
depletion than in the sf-FM phase.
Also, the slight occupation of $d_{z^2}$ states on the subsurface atom is
larger for the reconstructed sf-AFM phase, but not sufficient to 
propagate the spin alignment further into the bulk.
The considerable difference with respect to the LaMnO$_3$ JTD is that
its distortion is extended, whereas at the surface the ordering is a local 
effect limited to the first two layers. 
This local spin-flip process is likely to be relevant to more general
situations in manganites, such as at surfaces and interfaces of doped
systems where it may affect spin transport, and at Mn sites neighboring
O vacancies, as in CaMnO$_{3-x}$.\cite{Zampieri}

To summarize, we have described a spin-flip process at the Mn-terminated
(001) surface of CaMnO$_3$ that is driven by symmetry lowering due to 
surface formation which causes the partial occupation of the 
$e_g$ $d_{z^2}$ surface states.  This partially occupied narrow $d_{z^2}$
band may display correlated electron behavior. 
This $d_{z^2}$ occupation reverses the magnetic alignment (from AFM to FM) 
at the surface in the direction orthogonal to the surface but conserves
the AFM symmetry along the surface. The surface states are almost completely 
polarized, but AFM symmetry requires that both spin states occur in equal
number, so this result may be difficult to verify experimentally. 

This research was supported by National Science Foundation grant
DMR-9802076.  Calculations were done at the Maui High Performance
Computing Center.


\begin{table}
\centering\begin{tabular}{ccccc}
   & PM & FM & A AFM & G AFM \\
\hline
E(eV)         & +0.56  & +0.16  & +0.08 & 0 \\
M ({$\mu_B$}) &  0     & 2.56   & 2.48   & 2.36   \\
\end{tabular}
\caption{Energies (per formula unit, referred to that of the stable G-type 
AFM phase) and magnetic moments on Mn atoms for the bulk CaMnO$_3$ in 
different magnetic phases.
\label{bulk}}
\end{table}

\begin{table}
\centering\begin{tabular}{cccc}
   & UNREC & SPIN-FLIP FM & SPIN-FLIP AFM \\
\hline
E(eV)          & 0       & +0.12  & --0.12  \\
$\Delta E$(eV) & 0.09    &        & 0.05    \\
 W(eV)         &  5.65   & 5.64   &  5.73 \\
\end{tabular}
\caption{Energies, per surface $\sqrt 2 \times \sqrt 2$ cell, of the 
(001) CaMnO$_3$ surface in different magnetic 
phases (see the text); $\Delta E$(eV) are the corresponding relaxation
energies, and W the workfunctions.
\label{tavola}}
\end{table}


\end{multicols}
\end{document}